\documentclass[twocolumn,aps,prl,showpacs,floatfix,superscriptaddress]{revtex4}
\usepackage{epsfig}
\usepackage{bm}
\usepackage{color}

\usepackage[latin1]{inputenc}
\newcommand{\beq}{\begin{equation}}
\newcommand{\eeq}{\end{equation}}
\newcommand{\bea}{\begin{eqnarray}}
\newcommand{\eea}{\end{eqnarray}}

\newcommand{\nue}{\nu_{\mbox{\tiny eff}}}
\newcommand{\epse}{\epsilon_{\mbox{\tiny eff}}}
\usepackage{amsmath}
\usepackage{amssymb}
\usepackage{bm}
\usepackage{color}
\usepackage{graphicx}
\newcommand{\bx}{{\bm x}}
\newcommand{\br}{{\bm r}}    
\newcommand{\bv}{{\bm v}}

\newcommand{\be}{\begin{equation}}
\newcommand{\ee}{\end{equation}}

\begin{document}
\title{Intermittency and universality in fully developed inviscid and
  weakly compressible turbulent flows}

\author{Roberto Benzi} \affiliation{Dipartimento Fisica and INFN,
  Universit\`a di ``Tor Vergata'', Via della Ricerca Scientifica 1,
  00133 Roma, Italy} \author{Luca Biferale} \affiliation{Dipartimento
  Fisica and INFN, Universit\`a di ``Tor Vergata'', Via della Ricerca
  Scientifica 1, 00133 Roma, Italy}

\author{Robert T. Fisher} \affiliation{Center for Astrophysical Thermonuclear Flashes, The University of Chicago, Chicago, IL 60637\\
  Department of Astronomy and Astrophysics, The University of Chicago,
  Chicago, IL 60637}

\author{Leo P. Kadanoff} \affiliation{Department of Physics, The
  University of Chicago, Chicago, IL 60637\\Department of Mathematics,
  The University of Chicago, Chicago, IL 60637}

\author{Donald Q. Lamb}\affiliation{Center for Astrophysical Thermonuclear Flashes, The University of Chicago, Chicago, IL 60637\\
  Department of Astronomy and Astrophysics, The University of Chicago,
  Chicago, IL 60637}

\author{Federico Toschi} \affiliation{Istituto per le Applicazioni del
  Calcolo CNR, Viale del Policlinico 137, 00161 Roma, Italy, and INFN,
  Sezione di Ferrara, Via G. Saragat 1, I-44100 Ferrara, Italy}

\date{\today}
\begin{abstract}
  We performed high resolution numerical simulations of homogenous and
  isotropic compressible turbulence, with an average 3D Mach number close
  to $0.3$. We study the statistical properties of intermittency for
  velocity, density and entropy.  For the velocity field, which is the
  primary  quantity that can be compared to the isotropic
  incompressible case, we find no statistical differences in its
  behavior in the inertial range due either to the slight
  compressibility or to the different dissipative mechanism.  For the
  density field, we find evidence of ``front-like'' structures,
  although no shocks are produced by the simulation.
\end{abstract}

\maketitle
Fully-developed three-dimensional turbulence is characterized by an
intermittent energy flux from large to small scales. For
incompressible and viscous turbulent flows, dissipation occurs close
to the Kolmogorov scale $\eta$. According to the Kolmogorov theory, the statistical
properties of turbulence are scale-invariant within the inertial range
$\eta \ll r \ll L_0$, where $L_0$ is the scale of energy
forcing. Intermittency spoils dimensional scale invariance and is the
origin of anomalous scaling \cite{frisch:1995}. There exist a number of
numerical and experimental results aimed at investigating the
intermittency in turbulence for incompressible and viscous flows. On
the other hand, few investigations have been reported so far for the
case of weakly compressible and inviscid turbulence, relevant to many
astrophysical and geophysical problems \cite{elmegreen:2004,
  kallistratova:2002}.  In this letter we present a high-resolution
numerical simulation of homogeneous and isotropic, three-dimensional
compressible and inviscid turbulence. Our main purpose is to
investigate intermittency in the inertial range and compare our
finding against known results for incompressible viscous
turbulence. For the case of driven and decaying supersonic turbulence
similar problems have been addressed in
\cite{porter:1998,porter:2002}.
% RTF : Define acronyms. Note the Flash center is not capitalized, but the
% code itself is. I refer to it as a 'simulation framework' here.
For this simulation, we use the FLASH 3 component-based simulation framework. 
While the FLASH framework was primarily designed to treat
compressible, reactive flows found in astrophysical environments
\cite{fryxell:2000}, it is generally applicable to many other types of
fluid phenomena. For this simulation, only the compressible
hydrodynamics module based on the higher-order Godunov Piecewise
Parabolic Method (PPM) was used
\cite{Colella:1984yq}. The algorithmic methodology of the FLASH
framework and the computer science aspects of this turbulence
simulation have been described in further detail elsewhere
\cite{Fisher:2007rt}.  The equation of motions solved are the Euler
fluid equations with forcing: \bea
\label{1}
\frac{\partial \rho}{\partial t} + {\bm \nabla} \cdot( \bv \rho) = 0 \\
\label{2}
\frac{\partial \rho \bv}{\partial t} + {\bm \nabla} \cdot (\bv \bv \rho) = - {\bm \nabla} P + {\bm F}\\
\label{3}
\frac{\partial \rho E}{\partial t} + {\bm \nabla} \cdot \left[\bv(\rho E +P)\right] = 0 \\
\label{4}
P = (\gamma-1) \rho U\,,\;\;\;\; E = U + \frac{1}{2}\rho v^2\;, \eea
where $\rho$ is the density, $\bv$ the velocity, $P$ the pressure, $E$
the total energy, $U$ the internal energy and $\gamma$ is the ratio of
the specific heats in the system, equation (\ref{4}) being the
equation of state for our system. The effect of the large scale
forcing ${\bm F}$ in (\ref{2}) gives rise to a turbulent flow whose
energy is transferred from scale $L_0$ towards small scales
\cite{Fisher:2007rt}.  The energy input $\int \bv {\bm F} \,d^3x $
produces an increase of the internal energy $U$, which grows in
time. One can easily show that the quantity $P\partial_i
v_i$ represents the energy transfer from kinetic to internal energy of
the flow, which acts primarily on the smallest scales.  The mean 
sound speed increases slightly in time as well,
though the 3-D RMS Mach number is roughly $0.3$ (1-D Mach number
$0.17$) throughout. The numerical simulation was done for isotropic
and homogeneous forcing \cite{eswaran:1988} with a state-of-the-art resolution of $1856^3$
grid points.  The integration in time was done for $3$ eddy turnover
times after an initial transient evolution beginning from rest.  A key
feature of the numerical simulation is that while the flow is formally
inviscid, some viscosity is introduced as a result of the numerical
scheme. The numerical treatment of turbulent flows used in the FLASH
simulation is sometimes referred to as an Implicit Large Eddy
Simulation (ILES), to be distinguished from a full Direct Numerical
Simulation (DNS) of the Navier-Stokes equations.  Thus, one might
expect that the dynamics of the flow may significantly differ from
incompressible and viscous high Reynolds number turbulence.

%%%%% FIGURE 3 %%%%%%%%%%%%%%%%%%
\begin{figure}
  \centering \epsfig{clip=,width=\hsize,file=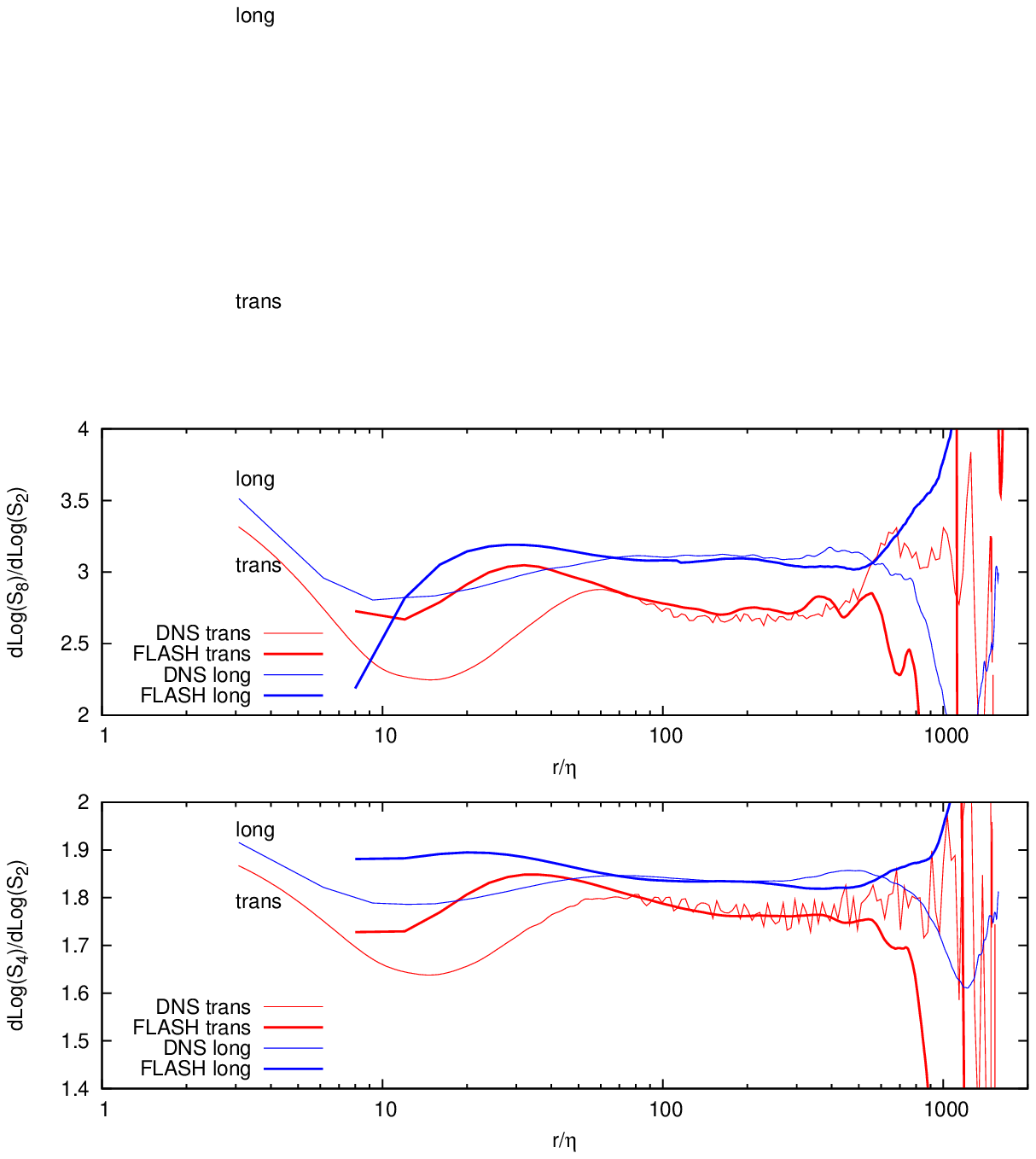}
  \caption{Local slopes of the 8th-order (top) and 4th-order (bottom) 
    longitudinal (blue lines) and transverse (red lines) structure
    functions as obtained from the present ILES (bold lines) and
    compared against a DNS of the Navier-Stokes equations
    \cite{gotoh:2002} at comparable Reynolds numbers (narrow 
    lines).}
  \label{fig3}
\end{figure}
%%%%%%%%%%%%%%%%%%%%%%%%%%%%%%%%%% 
%%%%% FIGURE 4 %%%%%%%%%%%%%%%%%%
%\begin{figure}[!b]
%  \centering \epsfig{width=\hsize,file=fig2.eps}
%  \caption{Comparison of the Kolmogoroff equations, with the effective
%    viscosity computed in the text, (black line) against the ILES
%    result of the third order structure functions (symbols).}
%\label{fig4}
%\end{figure}
%%%%%%%%%%%%%%%%%%%%%%%%%%%%%%%%%% 

%%DQL : Introduced new paragraph here to highlight key question.
A key question is therefore whether the statistical properties of the
inertial range are significantly different from to the homogeneous
and isotropic turbulence observed in Navier-Stokes.  To answer this
question, we employ the concept of anomalous exponents in measuring
the statistical properties of velocity fluctuations in the inertial
range. In particular, we use the well known method of Extended
Self-Similarity (ESS) \cite{benzi:1993}, which allows us to accurately
estimate the anomalous exponents of the longitudinal $S_P^{(L)} \equiv
\langle [\delta \bv(\br) \cdot {\hat \br}]^p \rangle$ (where $\delta
\bv(\br) = \bv(\bx+\br)-\bv(\bx)$ and $\langle ... \rangle$ means
average over the volume and in time) and transverse structure
functions $ S_P^{(T)} \equiv \langle [\delta {\bv}({\br}_T) ]^p
\rangle $ (where ${\br}_T \cdot {\bv} = 0$). We denote the
corresponding scaling exponents by $\zeta_p(L)$ and $\zeta_p(T)$. Our
numerical result for $\zeta_p(L),\zeta_p(T)$ agrees remarkably well with previous
numerical and experimental data. This is shown in figure \ref{fig3},
which compares the ESS local slope $ d \log(S_P) / d\log(S_2)$, for
both longitudinal and transverse structure functions $p=4$ and $p=8$
with the DNS simulations performed for incompressible Navier-Stokes
equation at comparable Reynolds numbers \cite{gotoh:2002}. Figure
\ref{fig3} shows several interesting features.  First, there exists a
range of scales ($r/\eta \ge 50$) where the local slope is almost
constant, i.e.  where we can detect accurately an anomalous scaling
exponent. Second, as one can see, in the inertial range our ILES
results give exactly the same anomalous scaling as that obtained for
the incompressible Navier-Stokes equation \cite{gotoh:2002}. There
is a clear difference between our results and the Navier-Stokes case
in the dissipation range ($r/\eta \le 20$), where the Navier-Stokes
solutions show a well defined ``dip'' effect (as qualitatively
predicted by the multifractal theory \cite{frisch:1991}), while ILES
behaves differently.  The different behavior in the dissipation range
is expected since the ILES does not dissipate energy in the
``standard'' Navier-Stokes way. On the other hand, the remarkable
agreement in the inertial range allows us to claim that the inertial
range properties are independent of the dissipation mechanism.  This
is an important statement, which has been questioned several times in
the past, and supports the conjecture that the statistical properties
of turbulence in the inertial range are universal and independent of
the dissipation mechanisms. This is one of our main results.

%%%%% FIGURE 5 %%%%%%%%%%%%%%%%%%
%\begin{figure}
%  \centering \epsfig{width=\hsize,file=fig3.eps}
%  \caption{Compensated structure functions of the density $D_6(r)
%    r^{-z_\infty}$ (solid line) and $D_5(r) r^{-z_\infty}$ (dashed
%    line). In the insert, we show the scaling exponents of the density
%    structure functions ($\bullet$) and the entropy structure
%    functions ($\blacktriangle$). Errors include both statistical
%    fluctuations and the uncertainty in the fit to the inertial
%    range.}
%\label{fig5}
%\end{figure}
\begin{figure}
  \centering \epsfig{width=.8\hsize,file=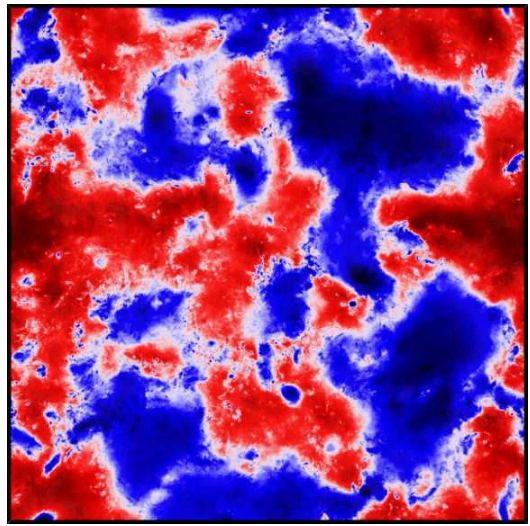}
  \caption{Two dimensional section of the density field $\rho$ at a
    given time: large regions with smooth density variations  are separated by sharp
    cliffs.}
  \label{fig1}
\end{figure}
%%%%%%%%%%%%%%%%%%%%%%%%%%%%%%%%%% 

%%DQL : Introduced new paragraph here to emphasize the answer to key question.
We also note that the difference between longitudinal and transverse
scaling exponents observed in homogeneous and isotropic DNS
\cite{gotoh:2002}, is also seen in our numerical results. This
discrepancy is an open theoretical issue, not explainable using
standard symmetry argument in homogeneous and isotropic turbulence
\cite{biferale:2005d}.  Whether this remains true at higher Reynolds
numbers is an open question (see also \cite{chen:1997} for a
discussion on this point).

%%RTF : Introduced new paragraph here to mark new thought.
Although the integration is formally inviscid, there is a net energy
transfer from the turbulent kinetic energy $1/2\rho v^2$ to the
internal energy. Thus we may consider that an effective viscosity
$\nue$ is acting on the system. In order to estimate $\nue$ we can
proceed as if the Kolmogorov equation --with effective parameters--
applies to our case: \be
\label{kolmo}
S_3^{(L)}(r) = -\frac{4}{5} \epse r + 6 \nue \frac{d}{dr}
S_2^{(L)}(r) \ee
%%%%% FIGURE 1 %%%%%%%%%%%%%%%%%%
\begin{figure}
  \centering \epsfig{width=\hsize,file=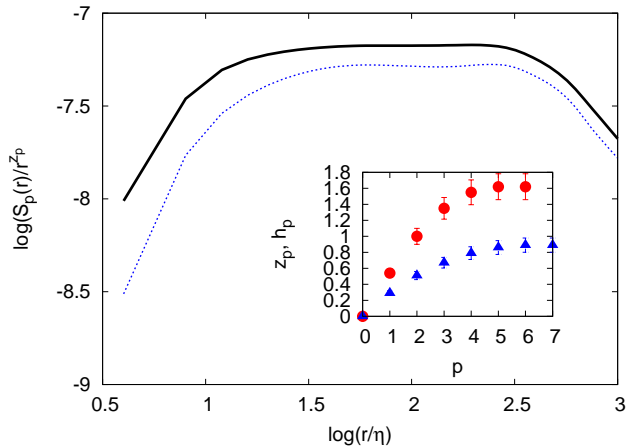}
  \caption{Compensated structure functions of the density $D_6(r)
    r^{-z_\infty}$ (solid line) and $D_5(r) r^{-z_\infty}$ (dotted
    line). In the insert, we show the scaling exponents of the density
    structure functions ($\bullet$) and the entropy structure
    functions ($\blacktriangle$). Errors include both statistical
    fluctuations and the uncertainty in the fit to the inertial
    range.}
\label{fig5}
\end{figure}
%%%%%%%%%%%%%%%%%%%%%%%%%%%%%%%%%% 
A fit of our data with this formula gives, $\epse=0.054, \nue
= 8.3 \cdot 10^{-6}$, which corresponds to a  Kolmogorov scale, $\eta
= (\nue^3 / \epse^{1/4})$,  equivalent to roughly half
grid cell and to $R_{\lambda} \sim 600$.
%let $\epsilon \equiv \langle \bv\cdot{\bm F} \rangle $ be the mean
%rate of energy dissipation, $\epsilon \sim 0.09$ in dimensionless
%units in which the box size, fluid density, and sound speed are all
%unity. We can also evaluate $\epsilon = 7.5 \nue \langle (\nabla v )^2
%\rangle$ and $\langle (\nabla v )^2 \rangle = \lim_{r\rightarrow
%  0}(1/2r) d S_2^{(L)}dr$, which gives $\nue = 5. 10^{-5}$. According
%to this value of $\nue$, the corresponding $Re_{\lambda} \equiv (7.5
%\langle (\nabla {v} )^2 \rangle^{1/2} v^2_{rms}) / \epsilon \sim 220$,
%where $v^2_{rms}$ is the mean square average velocity; i.e., $
%2v^2_{rms} = \lim_{r \rightarrow \infty} S_2^{(L)}(r)$.  The
%Kolmogorov scale $\eta = (\nue^3 / \epsilon)^{1/4}$ is equivalent to 5
%grid cells.
%A check of the accuracy in estimating $\nue$ can
%be done by using (\ref{kolmo}). In particular, we can employ ESS to
%obtain an accurate fit of $S_2^{(L)}$ as a function of $S_3^{(L)}$:
%i.e., $S_2^{(L)} = A [S_3^{(L)}]^{\zeta_2}$, which gives $\zeta_2
%=0.71$ and $A \sim 0.88$. Then we can solve (\ref{kolmo}) for
%$S_3^{(L)}$, given $\nue$, and compare it against the value of
%$S_3^{(L)}$ obtained by the ILES. This comparison is done in figure
%\ref{fig4} where the symbols refer to ILES results and the line
%corresponds to the numerical solution of (\ref{kolmo}). As one can see
%the agreement is rather good, supporting the estimate of $\nue$
%previously done. This result agrees with our previous estimate of the
%dissipation range (see figure \ref{fig3}).
 The dynamical effects of the
effective viscosity are however different from what one usually
observes in the Navier-Stokes equations; i.e., the dissipation range
does not behave the same as in the Navier-Stokes solutions. Thus,
while we still observe an effective dissipation in the energy flux
(i.e., in the third-order structure functions), the behavior in the
dissipation range is different because the dissipation mechanism in our
simulation is different:  the increase of internal energy is due to
viscosity resulting from the numerical
scheme.

%%%%% FIGURE 2 %%%%%%%%%%%%%%%%%%
%\begin{figure}
%  \centering \epsfig{width=.8\hsize,file=fig5.eps}
%  \caption{Same as in figure \ref{fig1} for the entropy field
%    $S=\log(P/\rho^{\gamma})$. }
%  \label{fig2}
%\end{figure}
%%%%%%%%%%%%%%%%%%%%%%%%%%%%%%%%%% 
In figure \ref{fig1} we show the density field in the system at a
given time. One can easily recognize the existence of large density
gradients due to compressibility. Another interesting quantity to look
at is the entropy $S$ defined as $S \equiv \log(P/\rho^{\gamma})$.
Using equations (\ref{1})-(\ref{4}) one can obtain the following
equation for the entropy: $\partial_t S + \bv \cdot \nabla S = 0$.
%\be
%\partial_t S + \bv \cdot \nabla S = 0
%\label{6}
%\ee
This equation tells us that $S$ satisfies an equation similar to the
case of a passive scalar advected by the velocity vector $\bv$,
although $S$ cannot be considered a passive quantity in this case. 
%In
%figure \ref{fig2} we show the enstropy field $S$ for the same time and
%slice as in figure \ref{fig1}, 
Strong variations are also detectable in the field $S$ (not shown).
Recently, the statistical properties of density fluctuations have been
investigated for supersonic turbulence characterized by large Mach
number. A rather large effect is caused by the formation of shock
waves and fronts \cite{porter:1998,porter:2002,Kritsuk:2007fj}. Here, the
3D Mach number is of order $0.3$ on average and, consequently, we should
expect that shock waves are not important (although the
compressibility degree may reach a maximum excursion where Mach $\sim
O(1)$).  Actually, it has been shown in experiments and direct
numerical simulations of passive scalars, that front-like structures
are frequently observed \cite{celani:2001,warhaft:2000a}. A front-like structure
on a quantity $Q$ is characterized by a ``local scaling'' property
$Q(x+r)-Q(x) \sim const$ for $x+r$ and $x$ selected on the two
different sides of the front. If these ``front-like'' structures play
a significant role in the statistical fluctuations of $Q$, one should
observe quite intermittent anomalous scaling for the structure
functions of $Q$. In particular the anomalous exponents
should approach a constant value for $p \rightarrow \infty$.

In order to study the statistical properties of the density and
entropy fluctuations, we can introduce the density structure functions
$D_p(r) = \langle [\delta \rho(r) ]^p \rangle$ and the entropy
structure functions $ E_p(r) = \langle [\delta S(r) ]^p \rangle$. The
ILES result shows that $D_P(r)$ and $E_p(r)$ are scaling functions of
$r$ in the inertial range; i.e., $D_p(r) \sim r^{z_p}$ and $E_p(r)
\sim r^{h_p}$.  The values of $z_p$ and $h_p$ are shown in the insert
of figure \ref{fig5} together with the compensated plot $D_5 r^{-z_5}$
and $D_6 r^{-z_6}$. From figure \ref{fig5} one can appreciate the
quality of the scaling in the inertial range. We remark that the
scaling exponents $z_p$ and $h_p$ show quite anomalous behavior. For
$p \le 3$ the values of $z_p$ are larger than the corresponding values
of $\zeta_p(L)$ and $\zeta_p(T)$ which means that density is somehow
smoother than the velocity field.  The anomalous exponents for both
the density and the entropy structure functions become constant at
large order. In particular, defining the saturation exponents to be
$z_{\infty}$ and $h_{\infty}$, we estimate $z_{\infty} = 1.62 \pm
0.10$ and $h_{\infty} = 1.00 \pm 0.10$. A similar analysis for the
pressure field (not shown) shows that the structure functions of the
pressure field $P$ have the same scaling exponents and of the same
saturation exponents as those for the density field.
%%%%% FIGURE 6 %%%%%%%%%%%%%%%%%%
\begin{figure}
  \centering \epsfig{width=\hsize,file=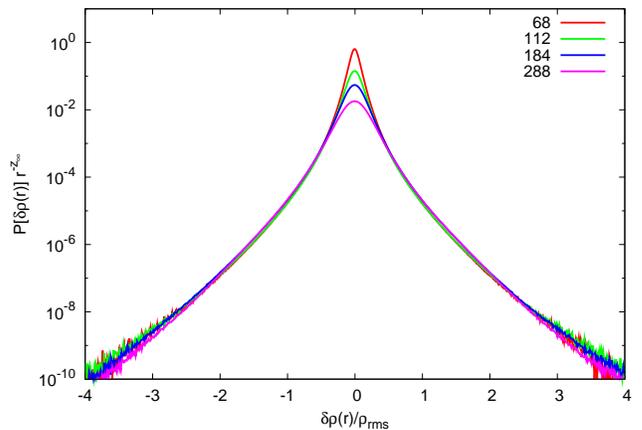}
  \caption{Rescaled probability distribution function, $P[\delta
    \rho(r)] r^{-z_{\infty}}$. Different curves corresponds to
    different values of $r/\eta$ in the inertial range (respectively, from
    top to bottom, $68, 112, 184, 288$.) }
\label{fig6}
\end{figure}
%%%%%%%%%%%%%%%%%%%%%%%%%%%%%%%%%% 
Using equation (\ref{1}) and assuming stationarity, one can derive an
exact relation for the ``energy'' flux of density fluctuations
$\langle (\delta \rho(r))^2 \delta v(r) \rangle$; namely, $\langle
(\delta \rho(r))^2 \delta v(r) \rangle = 2 \langle (\rho(x+r))^2 {\bm
  \nabla} \cdot {\bv}(x) \rangle$.  The above equation is fundamental
in relating the scaling of the density fluctuations in the inertial range to
a flux-like quantity  of ``density energy'' analogous to the 
the case of a passive or active scalar.  The single point value
$\langle (\rho(x))^2 {\bm \nabla} \cdot {\bv}(x) \rangle$ should play
the same role as ``energy dissipation'' does in the $4/5$ law for the
energy cascade in turbulence.  This issue deserves more detailed study
in the future.

According to the multifractal theory of turbulence, the effect of
saturation in the exponents $z_p$ and $h_p$ is equivalent to saying that the
tail of the probability distribution $P[\delta \rho(r)]$ should behave
as $\sim r^{z_{\infty}}$ for any $r$.  Thus we should expect that the
functions $r^{-z_{\infty}}P[\delta \rho(r)]$ should collapse on the
same ``universal'' distribution for $r$ in the inertial range
\cite{celani:2001}. In figure \ref{fig6} we show that this is exactly
the case for ILES result. Let us note that, as shown in figure
\ref{fig5}, $z_{\infty}$ is larger than $h_{\infty}$. Using the
multifractal theory, one can relate the saturation exponent to the
fractal dimension of the ``front-like'' structures; i.e., $D_{\rho} =
3- z_{\infty}$ and $D_S = 3 - h_{\infty}$, where $D_{\rho}$ and $D_S$
are the fractal dimensions of the ``front-like'' structures for the
density and entropy, respectively.  Our findings show that $D_S$ is
larger than $D_{\rho}$; i.e., fronts in the entropy field are easier
to form than those in the density field. This result may not be
surprising if we observe that large entropy fluctuations are produced
by large pressure fluctuations {\it or} by large density
fluctuations.  Thus the fractal dimension of the entropy fronts may
be larger than those of density and pressure. The above argument
implies that entropy is a more intermittent quantity than are density
and pressure. The fact that the entropy
satisfies a transport equation tells us it is a conserved quantity
along Lagrangian trajectories of the numerical simulations.  This
could suggest some connection between the existence of front-like
structures and the behavior of inertial particles in {\it
  incompressible} turbulence, where it is known that particles tend to
form multifractal sets with correlation dimension as low as $2$
\cite{bec:2007,eaton:1994}. We feel that important results can
therefore be achieved by building a systematic ``bridge of knowledge''
between Lagrangian dynamics in compressible flows and entropy
statistics in compressible flows.

Let us summarize our results. Using a numerical simulation of inviscid
homogeneous, isotropic weakly-compressible turbulence, we find that
the scaling properties of the velocity field in the inertial range are
in excellent agreement with those observed in DNS of the Navier-Stokes
equations \cite{gotoh:2002}. This result supports the statement that
the nature of the dissipation does not affect the statistical
properties of the inertial range; i.e., turbulence is universal with
respect to the dissipation mechanism. We confirm that transverse and
longitudinal structure functions show  different scaling properties
(up to this Reynolds number).  We have also shown that, although
almost no shock waves are produced in the simulation, the density
fluctuations are characterized by ``front-like'' structures that
determine the tail of the probability distribution of $\delta
\rho(r)$.  Accordingly, the scaling exponents of density and entropy
structure functions, $D_p$ and $E_p$ saturate at large $p$.

The presence of front-like structures is well known in passive scalars
advected by incompressible turbulent flows
\cite{celani:2001,warhaft:2000a}, which suggests that the density
field behaves as a passive-like scalar at small Mach numbers. The
values of the saturation exponents for the density structure functions
are different from those observed in the 3D passive scalar case.  This
provides further confirmation of the fact that the scaling properties
of passive or ``passive-like'' quantities are not universal with
respect to the properties of the advecting velocity field
\cite{gawedzki:1995} (here the correlation between density and the
velocity is certainly different from that observed for true passive
scalar fields). The same argument may explain the difference between
the scaling properties of the entropy we see in our simulation and
those seen at larger Mach numbers in \cite{porter:1998,porter:2002}.

{\bf Acknowledgement} RB, LB and FT thank FLASH Center at University
of Chicago.  This research was supported in part by the DOE
ASC/Alliance Center for Astrophysical Thermonuclear Flashes at the
University of Chicago, Contract \#B523820.  The authors thank
K. Antypas, A. Dubey, D. Sheeler and the code group for crucial help
in performing the simulations and A. Apsden, P. Constantin, T. Plewa, and P. Rich
for insightful discussions.  The authors gratefully acknowledge the
use of the DOE ASC program's BG/L machine at LLNL.  The authors thank
T. Gotoh for providing data from DNS shown in fig.(\ref{fig3}) (see
also the iCFDdatabase \cite{iCFDdatabase}).

\bibliographystyle{apsrev}
\bibliography{toschi}
\end{document}